\documentclass[preprint]{aastex62}
\usepackage{amsmath}
\usepackage{natbib}
\usepackage{epstopdf}
\usepackage{float}
\usepackage[caption = false]{subfig}
\usepackage{graphicx,array,multirow}
\usepackage{listings}
\usepackage{rotating}

\newcommand{\lapprox} {\, \lower3pt\hbox{$\sim$}\llap{\raise2pt\hbox{$<$}}\,}
\newcommand{\gapprox} {\, \lower3pt\hbox{$\sim$}\llap{\raise2pt\hbox{$>$}}\,}
\newcommand{\eps}{\epsilon}

\begin{document}

\title{Determination of the total accelerated electron rate and power using solar flare hard X-ray spectra}

	\author[0000-0002-8078-0902]{Eduard P. Kontar}
	
	\author[0000-0001-6583-1989]{Natasha L. S. Jeffrey}
	\affil{School of Physics \& Astronomy, University of Glasgow, G12 8QQ, Glasgow, UK}

	\author[0000-0001-8720-0723]{A. Gordon Emslie}
	\affil{Department of Physics \& Astronomy, Western Kentucky University, Bowling Green, KY 42101}

\begin{abstract}

Solar flare hard X-ray spectroscopy serves as a key diagnostic of the accelerated electron spectrum. However, the standard approach using the collisional cold thick-target model poorly constrains the lower-energy part of the accelerated electron spectrum, and hence the overall energetics of the accelerated electrons are typically constrained only to within one or two orders of magnitude. Here we develop and apply a physically self-consistent warm-target approach which involves the use of both hard X-ray spectroscopy and imaging data. The approach allows an accurate determination of the electron distribution low-energy cutoff, and hence the electron acceleration rate and the contribution of accelerated electrons to the total energy released, by constraining the coronal plasma parameters. Using a solar flare observed in X-rays by the {\em RHESSI} spacecraft, we demonstrate that using the standard cold-target methodology, the low-energy cutoff (and hence the energy content in electrons) is essentially undetermined.  However, the warm-target methodology can determine the low-energy electron cutoff with $\sim$7\% uncertainty at the $3\sigma$ level and hence permits an accurate quantitative study of the importance of accelerated electrons in solar flare energetics.
\end{abstract}

\section{INTRODUCTION}\label{introduction}

Over the last 50 years or so, solar flares have been observed over a very broad range of frequencies, from radio to gamma-rays. Hard X-ray (HXR) emission remains the key diagnostic to quantitatively determine the properties of accelerated electrons \citep[see, e.g.,][for  reviews]{2008LRSP....5....1B,2011SSRv..159..107H} and provides the essential properties (e.g., magnitude, shape) of the electron distribution \citep[e.g.,][]{2016ApJ...830...28P}. This is because (i) HXR production is linearly related to the emitting electron spectrum, thus providing a relatively simple expression linking the electron and HXR spectra \citep[see, e.g.,][for  a review]{2011SSRv..159..301K}; and (ii) unlike radio waves or optical emission, the HXRs produced throughout the entire flare volume (including both the corona and chromosphere) are very weakly affected by propagation effects\footnote{the HXR albedo patch \citep{1977ApJ...215..666L,1978ApJ...219..705B,2006A&A...446.1157K} is the only component of flare radiation that is related to the (Compton) scattering of HXRs.}, so there is a very close relation between the radiation produced and the radiation observed.

Imaging spectroscopy observations of X-ray coronal sources and chromospheric footpoints \citep[see e.g.,][]{2003ApJ...595L.107E,2008A&A...489L..57K,2010ApJ...721.1933S}, using data from the Reuven Ramaty High Energy Solar Spectroscopic Imager {\it (RHESSI)} \citep{2002SoPh..210....3L}, have remarkably confirmed the validity of the so-called ``thick-target'' model \citep[e.g.,][]{1971SoPh...18..489B,1972SvA....16..273S}.  In this model, electrons are accelerated in the relatively tenuous corona and propagate downwards to the chromosphere, emitting electron-ion bremsstrahlung as they proceed, particularly at the dense chromospheric footpoints \citep[see, e.g.,][for a review]{2011SSRv..159..107H}. Using hard X-ray emission as a diagnostic of solar flare electrons requires knowledge not only of the bremsstrahlung cross-section but also of the energy evolution of the bremsstrahlung-producing electrons \citep[see, e.g.,][]{2003ApJ...595L.115B}. In general, the energy evolution of accelerated electrons is determined not only by binary Coulomb collisions with ambient electrons (and, to a lesser extent, ions), but also through scattering by plasma turbulence \citep[e.g.,][]{2017ApJ...835..262B} and collective effects such as return current Ohmic losses \citep[e.g.,][]{1977ApJ...218..306K,1980ApJ...235.1055E,2006ApJ...651..553Z,2017ApJ...851...78A} and Langmuir wave generation in the inhomogeneous plasma \citep[e.g.,][]{2012A&A...539A..43K}.

In a thick-target model, the electrons completely stop in the bremsstrahlung target and consequently the relationship between the magnitudes and shapes of the electron and HXR spectra depends only on the (energy-dependent) rate of energy loss of individual electrons within the target. For energies $E \gg k_B T$ (where $k_B$ is the Boltzmann constant and $T$ the temperature), the target is said to be ``cold'' (Section~\ref{cold-target-results}), and this has been the basis for much of the modeling of the HXR and electron spectral relationship in solar flares \citep[e.g.,][]{1971SoPh...18..489B,1972SvA....16..273S,1976SoPh...50..153L}. However, \cite{2003ApJ...595L.119E} has pointed out that the energy loss rate drops rapidly from its cold-target value to a value near zero as the accelerated electron energy approaches a few $k_B T$, and that recognition of this effect can significantly modify the number of electrons necessary to produce a given HXR burst (see Section~\ref{warm-target-results}). \citet{2005A&A...438.1107G} have further noted the importance of the collisional energy diffusion
of the few $k_B T$ electrons. However, all previous treatments have not considered the  transport of thermalized electrons. Recently, \citet{2015ApJ...809...35K} have developed a ``warm-target approach'' and demonstrated that these near-thermal electrons serve as an important constraint on the overall accelerated electron spectrum, and they also pointed out the need to take into account their spatial diffusive motion.

Since accelerated electron spectra are typically rather steep power-laws with spectral indices of $\delta \simeq 4$ or larger \citep[e.g.,][]{1988SoPh..118...49D,2002ApJ...569..459P}, the energy content in accelerated electrons is strongly dependent on the value of the lowest electron energy in the distribution, the low-energy cutoff $E_c$; underestimating the value of $E_c$ by even a factor of two can result in order-of-magnitude or greater overestimates of the energy content in accelerated electrons. Applying the cold-target model (Section~\ref{cold-target-results}) to observed solar flare HXR spectra strongly suggests that accelerated electrons account for a considerable fraction of the total magnetic energy released in the flare \citep{2004JGRA..10910104E,2005JGRA..11011103E,2012ApJ...759...71E}. However, because the lower energy end of the HXR spectrum is dominated by thermal bremsstrahlung, previous works \citep[e.g.,][]{2003ApJ...595L..97H,2012ApJ...759...71E} could not determine $E_c$ but provide only an upper limit and hence a lower limit to the energy contained in nonthermal electrons. In an attempt to provide a better estimate of the energy content in accelerated electrons, \cite{2016ApJ...832...27A,2017ApJ...836...17A}
included an approximate \citep[][see Section~\ref{warm-target-results} below]{2015ApJ...809...35K} treatment of warm-target effects in their analysis of the relationship between the HXR and accelerated electron spectra. Their analysis led to the inference of very significant energy contents in accelerated electrons, and, for some events, produced the particularly intriguing result that the inferred energy content in accelerated electrons was larger than (and in some cases over an order of magnitude greater than) the estimate of the magnetic energy available. Therefore, despite these numerous efforts, the details of how the energy is partitioned in solar flares remains essentially unknown.

In this paper, we present (Section~\ref{warm-target-results}) a self-consistent warm-target algorithm to correctly and accurately calculate a value of $E_c$ for a given hard X-ray spectrum and for given thermal properties of the corona. The procedure, which involves the use of both X-ray spectroscopy and imaging data, allows an accurate determination of the electron spectrum at both high (nonthermal) and low (thermal) energies, and hence an accurate evaluation of the contribution of electrons to the overall energetics of a flare (Section~\ref{fun_method}). We discuss the results and their implications in Section~\ref{discussion}.

\section{Accelerated electrons in a cold thick-target}\label{cold-target-results}

The spatially-integrated HXR spectrum (photons~cm$^{-2}$~s$^{-1}$~keV$^{-1}$) observed at the Earth is given by

\begin{equation}\label{eq:tt_hxr}
I(\eps)=\frac{1}{4\pi R^2}\int_{\eps}^{\infty} Q(\eps,E) \, \langle nVF(E)\rangle \, dE \,\,\, ,
\end{equation}
where $\eps$ (keV) is the photon energy, $E$ (keV) is the electron kinetic energy, $R$ (cm) is the distance from the Sun to the Earth, and $Q(\eps,E)$ (cm$^2$~keV$^{-1}$) is the bremsstrahlung cross-section, differential in electron energy $E$. The quantity $\langle nVF(E)\rangle$ (electrons~cm$^{-2}$~s$^{-1}$~keV$^{-1}$) is the density-weighted mean electron flux spectrum, integrated over the flaring region \citep{2003ApJ...595L.115B}, and it should be noted that it is determined solely from observations of $I(\eps)$ and knowledge of the bremsstrahlung cross-section $Q(\eps,E)$. However, to proceed further to the determination of the {\it injected} (or accelerated) electron spectrum ${\dot N}(E_0)$ (electrons~s$^{-1}$~keV$^{-1}$) from the mean electron flux spectrum $\langle nVF(E)\rangle$ requires a model for the electron dynamics in the flare.  This is often based on the simple collisional cold thick-target model \citep{1971SoPh...18..489B}, which assumes that the electrons lose all their kinetic energy in Coulomb collisions with much less energetic electrons in the ambient plasma. For such a model, the electron energy loss rate is $dE/dt = - K n v/E$, where $n$ (cm$^{-3}$) is the ambient density, $v$ (cm~s$^{-1}$) the electron velocity, and $K=2\pi e^4 \ln \Lambda$ (cm$^2$~keV$^2$), with $e$ (esu) the electronic charge and $\ln \Lambda$ the Coulomb logarithm.  It follows  \citep[see][]{2003ApJ...595L.115B} that the relationship between the mean electron flux spectrum and the injected
rate spectrum ${\dot N}(E_0)$ (electrons~s$^{-1}$~keV$^{-1}$) is

\begin{equation}\label{eq:tt_fbar}
\langle nVF(E)\rangle = \frac{E}{K}\int_{E}^{\infty}\dot{N}(E_0) \, dE_0 \, ; \qquad \dot{N}(E_0) = - K \left [ \frac{d}{dE} \left ( \frac{\langle nVF(E)\rangle}{E} \right ) \right ]_{E=E_0} \,\,\, .
\end{equation}

Application of expressions~(\ref{eq:tt_hxr}) and~(\ref{eq:tt_fbar}) to the power-law form $I(\epsilon) \propto \epsilon^{-\gamma}$, that is often \citep[e.g.,][]{2011SSRv..159..107H} an excellent approximation to observed HXR spectra, shows \citep[e.g.][]{1971SoPh...18..489B} that the accelerated electron spectrum is also a power-law

\begin{equation}\label{eq:tt_ndot}
\dot{N}(E) = \dot{N}_0 \, \frac{\delta-1}{E_c} \left(\frac{E}{E_c}\right)^{-\delta} \,\,\, ,
\end{equation}
with $\delta = \gamma+1$.  Here $E_c$ (keV) is formally an arbitrary reference energy, and in practice it is taken to be a low-energy cutoff in the injected electron spectrum.  $\dot{N_0}$ (s$^{-1}$) is the total rate of electron acceleration $\dot{N_0}=\int _{E_c}^{\infty} \dot{N}(E_0) \, dE_0$. The associated total power $P$ (keV~s$^{-1}$) in the nonthermal electrons is

\begin{equation}\label{eq:tt_power}
P = \int_{E_c}^\infty E_0 \, N(E_0) \, dE_0 = \left ( \frac{\delta-1}{\delta-2} \right ) \, \dot{N_0} \, E_c = \frac{1}{\delta-2} \, E_c^{2-\delta} \, \left [ {\dot N}(E_c) \, E_c^\delta \right ] \,\,\, ,
\end{equation}
so that for a prescribed accelerated spectrum ${\dot N}(E_0)$, $P$ is strongly dependent on the value of the low-energy cutoff $E_c$.

A typical solar flare X-ray spectrum also contains very substantial thermal bremsstrahlung emission at energies $\eps \approx (5-30)$~keV. This emission, produced in plasma heated to temperatures $k_B T \simeq 2$~keV \citep[e.g.,][]{1980ApJ...239..725D,2004ApJ...605..921P,2014ApJ...787..122S}, effectively masks the bremsstrahlung produced by the accelerated electrons at lower energies, making the determination of $E_c$, and so $P$, a significant challenge. It also means that while the cold thick-target model may be quite valid in the relatively cool chromosphere, it incorrectly describes the evolution of electrons in the hotter corona, especially electrons with energies up to a few times the thermal energy $k_BT$ \citep{2003ApJ...595L.119E,2005A&A...438.1107G,2014ApJ...787...86J,2015ApJ...809...35K}.  As we have noted, because of the steeply decreasing form of the accelerated electron spectrum, it is these very electrons that carry the bulk of the energy.  Therefore, to correctly model the electron spectrum in this critical range we require the  warm-target equivalent of Equation~(\ref{eq:tt_fbar}), which should be consistent with the form of the X-ray spectrum at both low (thermal) and high (nonthermal) energies. We consider just such a model in Section~\ref{warm-target-results} below.

To demonstrate the extent to which the collisional cold thick-target model {\em cannot} accurately constrain the parameters of the accelerated electron distribution,
we performed a typical spectral fit to the {\em RHESSI} X-ray spectrum of flare SOL2013-05-13T02:12 occurring on 2013~May~13, using the Object Spectral Executive \citep[OSPEX;][]{2002SoPh..210..165S}. We created the count spectrum using the combined {\em RHESSI} front detectors \citep[excluding the inadequately operating Detectors~2 and~7;][]{2002SoPh..210...33S,2011SSRv..159..107H}, during the impulsive phase period from 02:09-02:10~UT. The count spectrum was fitted between the energies of 10-100~keV with the following components: an isothermal fit (using the function f\_vth, that accounts for the low-energy ($<30$~keV) thermal bremsstrahlung), a cold thick-target model (using the function f\_thick2, that accounts for the nonthermal bremsstrahlung produced by the accelerated electrons), and auxiliary functions representing a pile-up correction\footnote{(f\_pileup\_mod) The pile-up component accounts for those photons arriving at the detector at nearly the same time, that are detected as a single count with an energy equal to the sum of the individual photon energies.}, and one Gaussian line at 10~keV that accounts for an instrumental feature. Since the flare is located at the limb where albedo effects are minimal, we do not add the albedo function. The spectral range chosen avoids the majority of other line and instrumental features present at lower energies, and also the dominant background at higher energies.  We fit the spectrum multiple times using different fixed values of $E_{c}$, ranging between 5 and 40~keV.

\begin{figure}[pht]\centering
\includegraphics[width=0.60\linewidth]{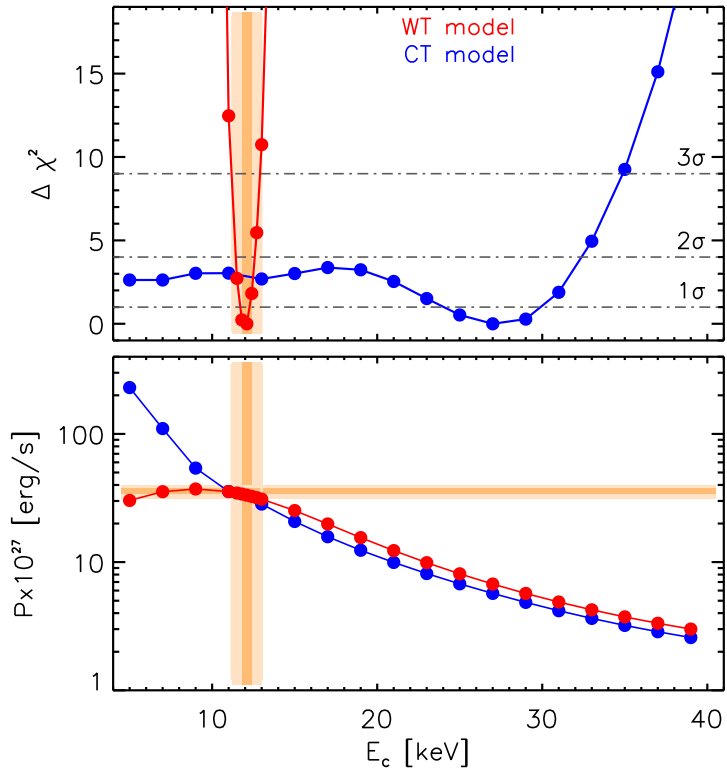}
\caption{Cold target (CT) and warm-target (WT) fitting of flare SOL2013-05-13T02:12, using a standard OSPEX approach with the parameters presented in Table \ref{tab1}. The {\em RHESSI} X-ray count spectrum, over the time interval of 02:09:00-02:10:00~UT, was fitted using the functions f\_vth+f\_thick2 (blue) and f\_vth+f\_thick\_warm (red). Other functions corresponding to pile-up + line (at 10~keV) were also used in the fitting process. {\em Top:} Full $\chi^{2}-\text{min}(\chi^{2})=\Delta \chi$ versus low-energy cutoff $E_{c}$. For the cold-target fit (CT: blue), a very large range of $E_{c}$ between 5-40~keV produces an acceptable fit to the spectrum at the $3\sigma$ level, resulting in a largely unconstrained value for $E_c$; however, the warm-target model (WT; red) discussed in Section~\ref{warm-target-results} provides the relatively well-constrained values of $E_{c} = (12.1 \pm 0.3)$~keV and $E_{c} = (12.1 \pm 0.9)$~keV at the $1 \sigma$ and $3\sigma$ levels, respectively. {\em Bottom:} nonthermal electron power, $P$, versus $E_{c}$. Since $E_{c}$ is not well constrained by the cold thick-target model, $P$ can range over some two orders of magnitude (blue).  The warm-target model (red) results
in constrained values of $P = (35.8 \pm 1.9) \times 10^{27}$~erg~s$^{-1}$ and $P = (35.8 \pm 4.4) \times 10^{27}$~erg~s$^{-1}$ at the $1\sigma$ and $3 \sigma$~levels, respectively. Dark orange: $1 \sigma$ level, light orange: $3 \sigma$ level.}
\label{fig1}
\end{figure}

The main results, displayed in Figure \ref{fig1}, show how the full $\chi^{2}$ \citep[see][for details]{1986nras.book.....P} and nonthermal electron power $P$ vary with $E_{c}$. The flat $\chi^2$ for the cold target in Figure~\ref{fig1} demonstrates that the low-energy cutoff $E_{c}$ cannot be determined at $2\sigma$-level. For the cold-target model, all values of $E_{c}$ below $\sim$33~keV are consistent (reduced $\chi^2 \sim 1-2$), with the observed X-ray spectrum at the $2\sigma$ level, and for this range of $E_c$, $P$ ranges over two orders of magnitude, from $P\approx 3 \times 10^{27}$~erg~s$^{-1}$ to $P\approx 3 \times 10^{29}$~erg~s$^{-1}$ (blue circles in Figure~\ref{fig1}). Hence the nonthermal electron power $P$ is very poorly constrained by application of the cold thick-target model. The lack of any substantial change in $\chi^{2}$ for the cold-target fit over a broad range of $E_c$ values (Figure~\ref{fig1}) is consistent with the detailed error analysis performed by  \citet{2013ApJ...769...89I}, who found that the probability density function for $E_c$ has a long asymmetric tail. As we shall show below, however, application of a self-consistent warm-target model results in a full $\chi^{2}$ that has a well defined range of acceptable $E_c$ values and hence of the nonthermal electron power $P$ (red points in Figure~\ref{fig1}). Using the warm-target approach defined below, the $\chi ^2$ analysis provides the low-energy cutoff $E_{c} = (12.1 \pm 0.3)$~keV and $E_{c} = (12.1 \pm 0.9)$~keV at the $1 \sigma$ (68\% confidence) and $3\sigma$ (99\%) levels, respectively.

\section{Warm-target considerations and the resolution
of the low-energy cutoff problem}\label{warm-target-results}

As they propagate through the 10-30~MK solar corona, electrons with energies of a few $k_BT$ suffer near-elastic collisions with ambient electrons of comparable energy and thus have an energy loss rate that is drastically lower than that in a cold-target approximation (see \cite{2003ApJ...595L.119E} and the simulations in \cite{2014ApJ...787...86J}). In a collisional cold thick-target model, the total number of accelerated electrons is not conserved; the electrons essentially ``disappear'' when they reach $E=0$. On the other hand, in a warm-target model the accelerated electrons retain a finite energy and hence their number {\it is} conserved. The continued injection of accelerated electrons into a target therefore systematically builds up the number of thermal electrons in that target, a process which ultimately saturates due to the diffusion of thermal electrons into the chromosphere (which, with a temperature $k_B T \lapprox 1$~eV, can accurately be considered a cold target). Therefore any consistent model of electron transport in collisional plasma must account for both the nonthermal and thermal components of the X-ray spectrum.

\citet{2015ApJ...809...35K} have shown that if electrons propagate in a region of warm plasma with length $L$ (cm), density $n$ (cm$^{-3}$) and temperature $T$ (K), the mean electron flux spectrum is related to the electron injection rate by (their Equation~(13)\footnote{note that the subsequent Equation~(24) in \citet{2015ApJ...809...35K} contains an erroneous upper limit of $\infty$ on the first integral. })

\begin{equation}\label{eq: fbar-solution-warm}
\langle nVF \rangle (E) = \frac{1}{2K} \, E \, e^{-E/k_BT} \, \int_{E_{\rm min}}^{E}
\frac{e^{E^\prime/kT} \, dE^\prime }{E^\prime \, G \left ( \sqrt{\frac{E^\prime}{k_BT}} \, \right ) } \,
\int_{E^\prime}^\infty \dot{N}(E_0) \, dE_0 \,\,\, ,
\end{equation}
where $G(x)=[{\rm erf}(x) - x \, {\rm erf}^{\prime}(x)]/2x^2$ and ${\rm erf}(x)$ is the error function. The limit $E_{\rm min}$ is given by

\begin{equation}\label{eq:E_min_adj}
E_{\rm min} \simeq 3 \, k_B T \, \left ( \frac{5 \lambda}{L} \right )^4 \,\,\, ,
\end{equation}
where $\lambda = (k_BT)^2/2Kn$ is the collisional mean free path, and it is determined by considering the warm plasma properties in the corona and by the gradual escape of electrons into the cold chromosphere --- see \citet{2015ApJ...809...35K} for details. In general, the shorter the electron mean free path, the smaller the value of $E_{\rm min}$ (Equation~(\ref{eq:E_min_adj})), and so the larger the number of electrons that will thermalize, and hence accumulate, in the corona (Equation~(\ref{eq: fbar-solution-warm})).

Equation~(\ref{eq: fbar-solution-warm}) replaces the cold thick-target result~(\ref{eq:tt_fbar}). The coronal parameters $T$, $n$ and $L$ that determine the value of $E_{\rm min}$, and hence the form of $\langle nVF \rangle (E)$, can best be obtained from a combination of X-ray spectroscopy and imaging observations: the distance between the coronal source and footpoint straightforwardly gives $L$, while the thermal fit to the HXR spectrum below $\sim$25~keV gives both the emission measure ${\rm EM} = n^2V$ and the temperature~$T$.

For loop lengths $L \simeq 10^9$~cm and coronal densities of order $10^{11}$~cm$^{-3}$ or larger, electrons at quite substantial energies up to $E \simeq 20$~keV are  thermalized in the corona, and, rather than becoming ``lost'' from the system (as they do in the cold thick-target formulation), they now make a significant, and observable, thermal contribution to the HXR spectrum.  This reduces the need for accelerated electrons in this energy range to create the HXR photons observed in this energy range and, as a result, the mean source electron spectrum $\langle nVF \rangle (E)$ needs to extend down only to fairly moderate energies $E$.  This effectively introduces a cutoff energy $E_c$ into the form of $\langle nVF \rangle (E)$, concomitantly reducing the required power in accelerated nonthermal electrons (cf. Equation~(\ref{eq:tt_fbar})). To quantify this effect, we write Equation~(\ref{eq: fbar-solution-warm}) in a simplified form, obtained by replacing the function $G(\sqrt{E^\prime/k_B T} ) \simeq \sqrt{E^\prime/\pi k_BT}$ (an approximation valid in the range $E \lapprox k_BT$), giving

\begin{eqnarray}\label{low-E-behavior}
\langle nVF \rangle (E) & \simeq & \frac{1}{2K} \, E \, e^{-E/k_BT} \, \sqrt{\pi k_B T} \int_{E_{\rm min}}^{E}
{E^\prime}^{-3/2} \, dE^\prime \,
\int_{E^\prime}^\infty \dot{N}(E_0) \, dE_0 \cr
& \simeq & \frac{1}{K} \, E \, e^{-E/k_BT} \, \sqrt{\pi k_B T} \, {E_{\rm min}}^{-1/2} \, {\dot N_0} \,\,\, ,
\end{eqnarray}
where we have used the fact that ${\dot N}(E_0)$ peaks at an energy $E_c$ substantially outside the $E \lapprox kT$ range of the approximation and have defined

\begin{equation}\label{n-dot-emin-def}
{\dot N_0} \simeq \int_{E_c}^\infty \dot{N}(E_0) \, dE_0\,\,\, .
\end{equation}
Then Equation (\ref{low-E-behavior}) can be rewritten in the Maxwellian form

\begin{equation}\label{eq:fbar_thermpart}
\langle nVF \rangle (E) = \Delta {\rm EM} \, \sqrt{\frac{8}{\pi m_e}} \, \frac{E}{(k_BT)^{3/2}} \, e^{-E/k_BT}
\end{equation}
where the emission measure

\begin{equation}\label{eq:delta_EM}
\Delta {\rm EM}  \simeq
\frac{\pi}{K} \, \sqrt{\frac{m_e}{8}} \, (k_B T)^2 \,\, \frac{\dot{N_0}}{E_{\rm min}^{1/2}} \,\,\, .
\end{equation}

$\Delta$EM quantifies the additional contribution to the overall inferred soft X-ray emission measure that results from the thermalization of accelerated electrons; it is a function of $E_{\rm min}$, a quantity that characterizes the fraction of the accelerated electron distribution that thermalizes in the hot coronal part of the loop. The value of $E_{\rm min}$ depends on the thermal collisional mean free path $\lambda$ like $T \, \lambda^4/L^4$ (Equation~(\ref{eq:E_min_adj})), so that $E_{\rm min}^{1/2} \sim T^{1/2} \lambda^2 / L^2 \sim T^{9/2}/n^2 \, L^2$ and thus $\Delta \mbox{EM}\propto \dot{N_0}n^2 \, L^2 \, T^{-5/2}$. Only electrons from the injected distribution with energy $E \lapprox \sqrt{2KnL}$ are essentially considered as thermalized and hence contribute to $\Delta EM$; electrons with energy in excess of this value can be treated as electrons interacting with a cold target in the conventional manner. Evidently, when the low-energy cutoff $E_c < \sqrt{2KnL}$, a substantial contribution $\Delta$EM is expected.

In the model of \cite{2015ApJ...809...35K}, the growth of emission measure in the thermal plasma due to the thermalization of freshly-injected low-energy energetic electrons in the coronal part of a flare loop is balanced with the diffusion of thermal electrons out of the hot coronal part of the loop into the chromosphere. This balance is achieved over a time corresponding to that for electron diffusion along the loop:

\begin{equation}\label{eq:t_diff}
\tau_\text{diff}\simeq \frac{L^2}{D} \,\,\, ,
\end{equation}
where $D={k_\text{B}T}\tau_e/m_e$ is the thermal diffusion coefficient and $\tau_e \simeq \sqrt{m_e} (k_B T)^{3/2}/\pi e^4 n \ln \Lambda $ is the thermal electron collision time. Using $n = 10^{11}$~cm$^{-3}$, $L = 10^9$~cm, and $T=20$~MK, one finds $\tau_e\simeq 10^{-2}$~s, $D \simeq 3 \times 10^{18}$~cm$^2$~s$^{-1}$ and $\tau_\text{diff}\simeq 30$~s. This diffusion timescale is comparable to the thermalization time for collisional \citep{1962pfig.book.....S} thermal conduction and it is sufficiently rapid that a steady-state balance can be achieved within the time associated with hard X-ray intensity fluctuations.
We also note (see above) that $\Delta EM \propto \dot{N_0} \, n^2 \, T^{-5/2}\propto \dot{N_0} \, n/D$; low diffusion coefficients lead to effective trapping of electrons and hence to a large emission measure from thermalized electrons.

For $E \gg k_B T$, Equation~(\ref{eq: fbar-solution-warm}) reduces to the cold thick-target form~(\ref{eq:tt_fbar}). Thus, we can approximate Equation~(\ref{eq: fbar-solution-warm}) as the sum of a ``thermal part'' and a ``nonthermal part'':

\begin{equation}\label{eq:fbar_warm2part}
\langle nVF \rangle (E) \simeq
\Delta EM \, \sqrt{\frac{8}{\pi m_e}} \, \frac{E}{(k_BT)^{3/2}} \, e^{-E/k_BT}+ \frac{E}{K}\int_{E}^\infty \dot{N}(E_0) \, dE_0 \,\,\, ,
\end{equation}
and it should be noted that when the low-energy cutoff in the injected distribution $E_c$ is larger than the energy $\sqrt{2KnL}$ that can be effectively stopped within the coronal part of the loop, the contribution $\Delta$EM of the thermal component becomes negligible and the cold-target approximation is recovered.

Equation~(\ref{eq:fbar_warm2part}), with a form of ${\dot N} (E_0)$ that assumes a power-law form at high energies, has been included in the Solar SoftWare (SSW) and OSPEX routines as the function f\_thick\_warm.pro\footnote{See OSPEX package \url{https://hesperia.gsfc.nasa.gov/ssw/packages/xray/idl/f_thick_warm.pro}}. The convolution of the mean source electron spectrum function $\langle nVF \rangle (E)$ obtained from Equation~(\ref{eq:fbar_warm2part}) with the bremsstrahlung cross-section $Q(\epsilon, E)$ \citep[e.g.,][]{1997A&A...326..417H} determines the hard X-ray flux, and by minimizing $\chi^2$, the best fit parameters can be found. The proper use of the expression~(\ref{eq:fbar_warm2part}) will be demonstrated in Section \ref{fun_method}. It is instructive to note that Equation~(\ref{eq:fbar_warm2part}) is a good approximation to Equation~(\ref{eq: fbar-solution-warm}) when $E_c<\sqrt{2KnL}$; this is readily seen from
Figures 3 through~5 in \citet{2015ApJ...809...35K}.

The above choice of a power-law form is, of course, not unique. For example, \citet{2014ApJ...796..142B} have demonstrated that the stochastic acceleration of electrons in the presence of both Coulomb collisions and velocity diffusion with a  coefficient that is inversely proportional to velocity leads to an accelerated electron distribution that  has the form of a kappa distribution. Although the total spatially integrated X-ray flux is not always consistent with a single kappa distribution, with an additional thermal component (presumably resulting from direct coronal heating associated with the magnetic reconnection process) required for an acceptable fit, kappa distributions have been found in some cases to be consistent with {\em RHESSI} observations of the overall X-ray spectrum \citep{2009A&A...497L..13K,2015ApJ...815...73B}. In Appendix \ref{kappa_section} we provide the formulae necessary to incorporate a kappa distribution as the nonthermal component of the accelerated electron spectrum.

\section{Application to Data}\label{fun_method}

Here we apply the warm-target method outlined in the previous section to flare SOL2013-05-13T02:12 to determine an acceptable form of ${\dot N}(E_0)$ and hence to infer the injected electron power~$P$. SOL2013-05-13T02:12 flare was one of the flares analyzed by \citet{2016ApJ...832...27A}, and they determined the nonthermal electron energy by using a simplified application of the warm-target model.

As noted above, the parameters for the nonthermal and thermal parts of the electron distribution are inter-related within the warm-target model.  Thus any algorithm aimed at producing a reliable estimate of the nonthermal electron power $P$ must also self-consistently determine the corresponding parameters (emission measure, temperature) of the thermal part of the spectrum. This is a key element of the method described below.  It should also be noted that the method requires the use of {\em both} hard X-ray {\em spectroscopy} and {\em imaging}. There are different ways to constrain the coronal thermal parameters. Below, we determine the thermal properties of the corona using the short time interval of 02:08:52-02:09:00~UT immediately preceding the time interval 02:09:00-02:10:00~UT used for the warm-target analysis. We stress that the following procedure is valid regardless of how the thermal properties of the corona, that will ultimately constrain the nonthermal electron energetics (see subsection \ref{plasma}), are determined.

\begin{figure*}[t]
\centering
\includegraphics[width=0.60\linewidth]{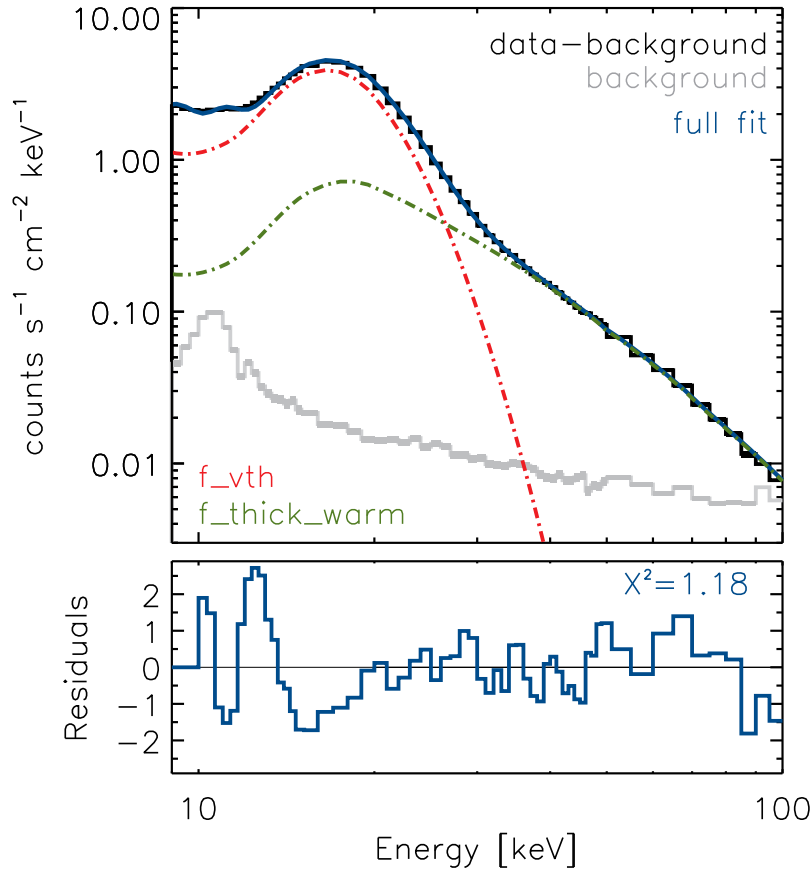}
\caption{Warm-target fitting of flare SOL2013-05-13T02:12. The {\em RHESSI} X-ray count spectrum minus the background (black) is integrated over a one minute time period from 02:09:00-02:10:00 UT. The {\em RHESSI} background is shown in grey. The primary fitting functions of f\_vth (isothermal, red) and f\_thick\_warm (warm-target, green) are shown; the fit parameters are listed in Table~\ref{tab1}. The count spectrum is fitted between the energies of 10-100~keV.  In this fit, the thermal plasma parameters of the corona, that constrain the nonthermal electron properties, are estimated at a preceding time interval of 02:08:52-02:09:00~UT, but the plasma parameters can be estimated using other time intervals. The full fit (blue) consists of f\_vth+f\_thick\_warm+pile\_up+line (at 10 keV). The fit residuals are plotted below the spectrum with a reduced $\chi^{2}=1.18$. The residuals show a poorer fit at 15-20~keV. This is likely due to the use of an isothermal approximation for the thermal component \citep[see][and the end of Section \ref{fun_method}]{2015A&A...584A..89J}.}
\label{fig 2}
\end{figure*}

\begin{figure*}[t]\centering
\includegraphics[width=0.4\linewidth]{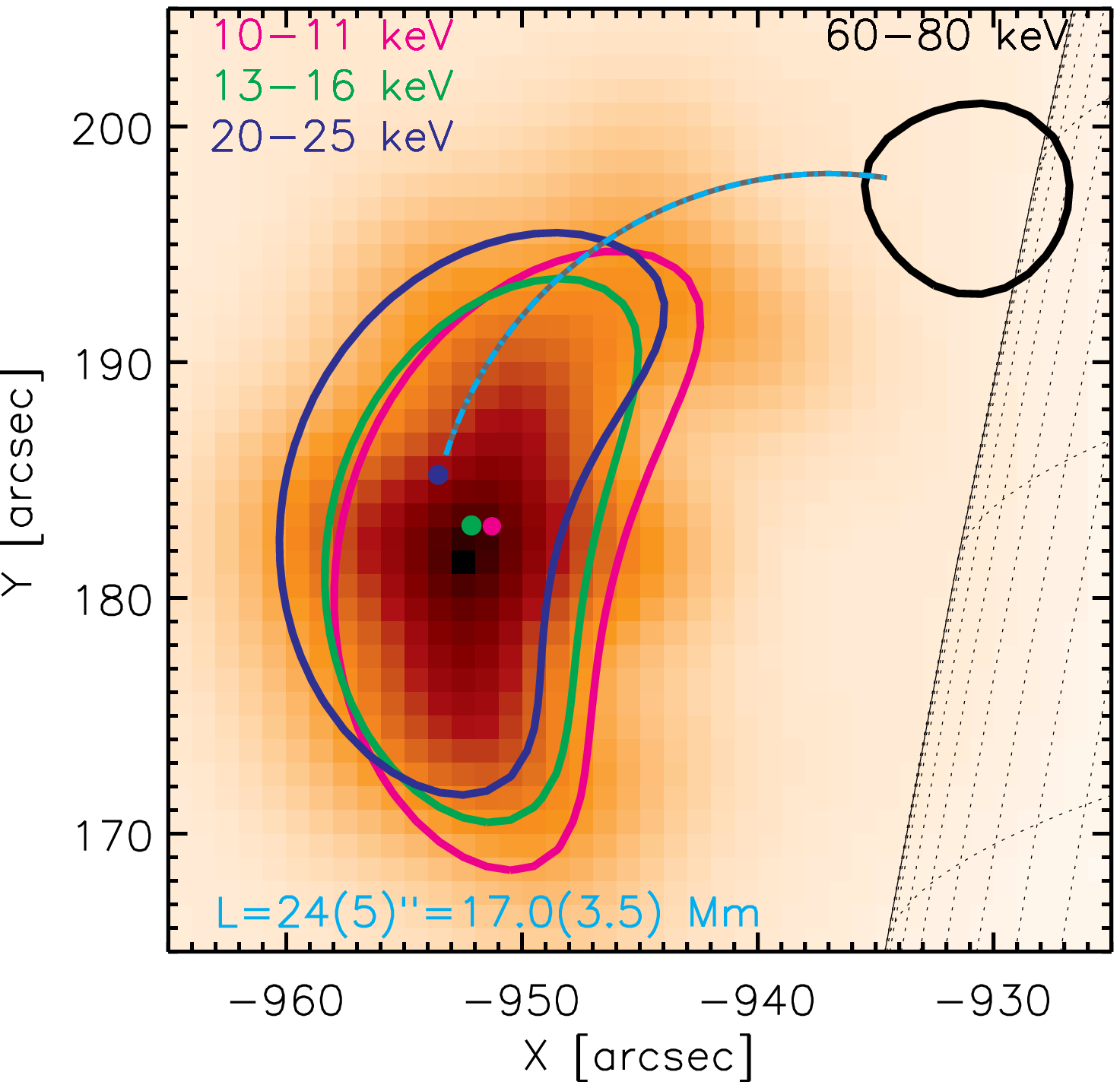}
\hspace{10 pt}
\includegraphics[width=0.4\linewidth]{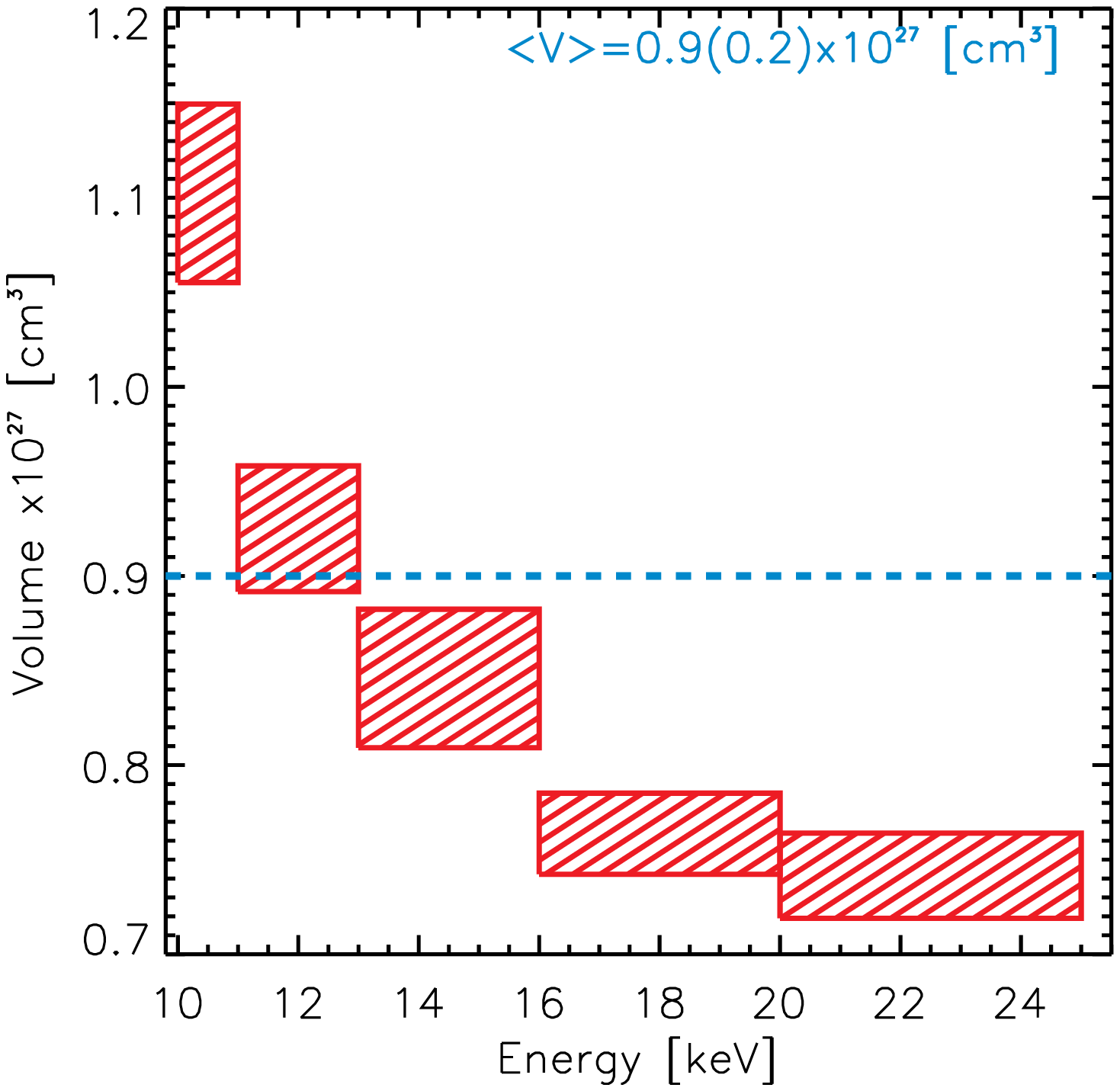}
\caption{X-ray imaging of flare SOL2013-05-13T02:12. {\em Left:} A CLEANed {\em RHESSI} X-ray image of SOL2013-05-13T02:12 integrated over a two minute interval from 02:08:00-02:10:00 UT (red-scale background image), over the energy range 13-16~keV. A coronal loop-top source is observed mainly at energies below 25~keV while a chromospheric footpoint source appears at higher energies (60-80~keV; this footpoint source is shown at the 50\% contour level). For this flare, the strong loop-top source was also imaged using the VIS\_FWDFIT algorithm to estimate the coronal source dimensions (loop-top length $l$ and width $w$) of the source by fitting a simple shape, such as an elliptical Gaussian, at different energy ranges. The volume of the loop-top source was then estimated using the cylindrical formula $V=\pi w^{2} l/4$.  The half loop length $L$ (the distance between the centre of the loop-top source and the chromospheric footpoint) is estimated by fitting an arc from the middle of the loop-top source and the footpoint, giving $L=24\arcsec \simeq 17$~Mm. An uncertainty of $\sim5\arcsec \simeq 3.5$~Mm is estimated. {\em Right:} Changes in the VIS\_FWDFIT volume with energy. We use the mean volume $\langle V\rangle=(0.86\pm0.20)\times10^{27}$~cm$^{3}\approx(0.9 \pm 0.2)\times10^{27}$~cm$^{3}$ (dashed blue line) to determine the parameter $n_{\rm loop}$ in f\_thick\_warm.}
\label{fig3}
\end{figure*}

\subsection{Description of the warm-target fitting procedure, and illustrative application}\label{method-description}

We now enunciate the general steps in the method and, by way of illustration, apply it to the flare SOL2013-05-13T02:12.  We have chosen to apply Equation (\ref{eq:fbar_warm2part}) with an assumed power-law form for $\dot{N}(E)$; use of other forms for $\dot{N}(E)$ (e.g, a kappa distribution) involves straightforward adjustments at the pertinent steps.

\begin{enumerate}

\item First we evaluate the plasma parameters in the flaring loop. We have chosen to evaluate the thermal parameters using the short time interval of 02:08:52-02:09:00~UT just before the studied interval of 02:09:00-02:10:00~UT. The HXR spectrum during the 02:08:52-02:09:00~UT interval was fitted with a combined thermal function plus a {\it cold} thick-target nonthermal function (e.g., using the SSW/OSPEX functions f\_vth + f\_thick2). Although both nonthermal and thermal components are included in the modeling, this step is primarily used to estimate\footnote{Here, we only estimate the plasma parameters using isothermal fitting.} the {\em thermal} plasma parameters of the coronal loop-top source, namely the temperature $T$ (keV) and the emission measure EM$_0$ (cm$^{-3}$); the nonthermal parameters will be determined more accurately later. The minimum value of the reduced $\chi^{2}$ gives the best estimates of $T$ and EM$_0$.  Even though the fitting is performed with an assumed power-law form for the cold thick-target nonthermal component, we have found that using another form (such as a kappa distribution) does not substantially affect the inferred values of the thermal parameters EM$_0$ and $T$.

   [ For SOL2013-05-13T02:12, the relative abundance parameter was fixed at the default value of unity, and we determined best-fit values of $T=(2.52\pm0.03)$~keV and EM$_{0}$=$(6.51 \pm 0.30) \times 10^{48}$~cm$^{-3}$  for 02:08:52-02:09:00 UT.    Note that throughout this work $1\sigma$ OSPEX uncertainties are used, (unless stated otherwise).~]

\item With the values of EM$_{0}$ and $T$ now determined, we fit the HXR spectrum during our study time interval of 02:09:00-02:10:00~UT. The HXR spectrum during this time interval is fitted with f\_vth and the warm-target fitting function f\_thick\_warm. Note that the total emission measure EM$=$EM$_{0}+\Delta$EM obtained by this fit includes any contribution $\Delta$EM from the thermalization of injected electrons. Any additional parameters/contributions pertinent to the observed flux (e.g., elemental abundances, albedo, pulse pile-up, lines, drm\_mod etc.) should also be included in this fit (Figure \ref{fig 2}). The function f\_thick\_warm contains 10 parameters: the first six parameters are exactly the same as for f\_thick2 (i.e., the electron acceleration rate $\dot{N_0}$ (s$^{-1}$), the spectral index (low) $\delta_{\rm low}$, the break energy $E_{B}$ (keV), the spectral index (high) $\delta_{\rm high}$, the low-energy cutoff $E_{c}$ (keV) and the high energy cutoff $E_{H}$ (keV)). The other four parameters relate to the properties of the thermal properties of the corona: the loop temperature $T_{\rm loop}$ (keV), loop density $n_{\rm loop}$ (cm$^{-3}$), and half-length of the loop $L$ (cm), and the relative elemental abundances.

\hskip 0.5in $\bullet$ As discussed in Step~1, an isothermal estimate of the loop-top temperature $T$ is determined using f\_vth during 02:08:50-02:09:00~UT, and we fix $T_{\rm loop} = T = 2.52$~keV.

\hskip 0.5in $\bullet$ To determine the density $n_{\rm loop}$, we use the estimate of EM$_{0}$ obtained in Step~1. The number density can then be estimated using $n_{\rm loop}=\sqrt{{\rm EM_{0}}/V}$, where $V$ is the volume of the emitting plasma, found from X-ray imaging (see Figure~\ref{fig3}).

\hskip 0.5in [ For SOL2013-05-13T02:12, the volume was estimated by applying the imaging algorithm Visibility Forward Fitting \citep[VIS\_FWDFIT;][]{2007SoPh..240..241S}. Using VIS\_FWDFIT, the volume of the coronal source varies with energy \citep[see][]{2015A&A...584A..89J}, so we determined the mean volume over the energy range of 10-25 keV (Figure~\ref{fig3}), yielding a value of $V = (0.86 \pm 0.20) \times 10^{27}$~cm$^3\approx(0.9 \pm 0.2) \times 10^{27}$~cm$^3$. Using the inferred values of EM$_{0}$ and $V$ (and their associated uncertainties) gives $n_{\rm loop}=(8.7 \pm 2.1) \times 10^{10}$ cm$^{-3}$. ]

\hskip 0.5in $\bullet$ The half length $L$ of the loop (the length from the coronal source to the chromospheric footpoint), is estimated directly from the image (see Figure~\ref{fig3}).

\hskip 0.5in [ For SOL2013-05-13T02:12, we estimate from the X-ray image (blue curve in Figure \ref{fig3} left) that $L=(24 \pm 5) \arcsec \sim (17 \pm 3.5)$~Mm. ]

\item Once the plasma parameters $n_{\rm loop}$, $T_{\rm loop}$ and $L$ have been determined and fixed, then the remaining nonthermal parameters of the f\_thick\_warm fit, namely ($\dot{N_0}$, $\delta_{\rm low}$, $E_{B}$, $\delta_{\rm high}$, $E_{c}$ and $E_{H}$), can be determined during 02:09:00-02:10:00~UT. As usual, we suggest that $E_{H}$ be fixed at the high default value and that $E_{B}$ and $\delta_{\rm high}$ are left free only if required (otherwise keep them fixed with $E_{B}$ above the highest energy in the fit).

\hskip 0.5in [ For SOL2013-05-13T02:12, we find $E_{c}=(12.1\pm0.3)$~keV, $\dot{N_0} = (12.8 \pm 0.6) \times 10^{35}$~electrons~s$^{-1}$ and $\delta_{\rm low}=\delta=(4.25\pm0.02)$ . ]

\item The nonthermal electron power can now be determined as $P=(\delta-1)/(\delta-2) \, \dot{N_0}E_{c}$, as for the cold thick-target case (Equation~(\ref{eq:tt_power})), since we assume a power-law form for the accelerated electrons.

\hskip 0.5in [ For SOL2013-05-13T02:12, we find $P=(35.8 \pm 1.9) \times 10^{27}$ erg~s$^{-1}$. ]

\end{enumerate}

All the major fit parameters for f\_vth and f\_thick\_warm for SOL2013-05-13T02:12 are shown in Table \ref{tab1}. Using the derived plasma parameters, we calculate a value of $\tau_{\rm diff}\sim30$~s, showing that the one minute observation time is well justified here.

\begin{table*}[pht]
\centering
\begin{tabular}[t]{|c|l|l|}\hline
~ & \multicolumn{1}{c|}{Parameter} & {\hfill Fit status\hfill} \\ \hline
\multirow{2}{*}{\rotatebox[origin=c]{90}{\parbox[c]{1.0cm}{\centering f\_vth}}} & EM$_{0}$ $= (6.51 \pm 0.30) \times10^{48}$~cm$^{-3}$ & Fixed, found from f\_vth at 02:08:52-02:09:00~UT.\\
& $T = (2.52 \pm 0.03)$~keV & Fixed, found from f\_vth at 02:08:52-02:09:00~UT.\\ \hline
\multirow{7}{*}{\rotatebox[origin=c]{90}{\parbox[c]{3.0cm}{\centering f\_thick\_warm}}} & $\dot{N_{0}} = (12.8 \pm 0.6) \times 10^{35}$~s$^{-1}$ & Free, found from warm-target fitting at 02:09:00-02:10:00~UT.\\
& $\delta_{\rm low} = 4.25 \pm 0.02$ & Free, found from warm-target fitting at 02:09:00-02:10:00~UT.\\
& $E_{c} = (12.1 \pm 0.3)$~keV & Free, found from warm-target fitting at 02:09:00-02:10:00~UT.\\
& $n_{\rm loop} = (8.7 \pm 2.1) \times 10^{10}$~cm$^{-3}$ & Fixed using EM$_{0}$ \& $V$ (from imaging at 02:08:00-02:10:00~UT).\\
& $T_{\rm loop} = (2.52 \pm 0.03)$~keV & Fixed and equal to $T$ (from f\_vth at 02:08:52-02:09:00~UT).\\
& $L = (17.0 \pm 3.5)$~Mm & Fixed from imaging at 02:08:00-02:10:00~UT.\\
& $P = (35.8 \pm 1.9) \times 10^{27}$~erg~s$^{-1}$ & Determined from $\dot{N_{0}}, \delta_{\rm low}, E_{c}$.\\ \hline
\end{tabular}
\caption{Best-fit parameters of f\_vth and f\_thick\_warm for the spectrum shown in Figure \ref{fig 2}. Only the f\_thick\_warm parameters related to the accelerated electron distribution (i.e., $\dot{N_{0}}, \delta, E_{c}$) are left free during warm-target fitting. The plasma parameters (i.e., EM$_{0}$, $T_{\rm loop}, n_{\rm loop}, L$) are determined using a combination of X-ray spectroscopy and imaging, and then fixed during fitting.}
\label{tab1}
\end{table*}

\subsection{Comparison with simple analytic estimates.}

A crude estimate of the cutoff energy $E_c$ can be obtained by considering the energy at which the systematic energy loss rate vanishes in the Fokker-Planck equation governing the evolution of the nonthermal electron spectrum. \citet{2015ApJ...809...35K} used this method to obtain the approximate result $E^*_c \simeq \delta \times k_{B}T$. Although this provides a useful, and easily applied, estimate of $E_c$ and hence of the electron power $P$ (Equation~(\ref{eq:tt_power})), it should be stressed that this simplified expression corresponds simply to the value at which the nonthermal component of $\langle n V F \rangle (E) = 0$; below this value of $E^*_c$ the value of $\langle n V F \rangle (E) $ is not zero, but in fact negative. Using $T_{\rm loop} \simeq 2.5$~keV and $\delta = 4.2$, the simple estimate $E^*_c = \delta \times T_{\rm loop}\;{\rm [keV]}$ gives $E^*_c \simeq 10$~keV; this was the approach used by \citet{2016ApJ...832...27A}. Here, $E^*_c$ is close to $E_{c}$ determined by the warm-target forward-fit method. However, this will not be true of all flares and hence we stress that the nonthermal electron parameters should only be determined using the warm-target forward-fit method described in Section \ref{method-description}.

The $E^*_c$ approximation was used by \citet{2016ApJ...832...27A,2017ApJ...836...17A} to compute the energy content in accelerated electrons.  In some cases, using the low-energy cutoff estimate $E^*_c \simeq \delta \times k_{B}T$ \citep{2015ApJ...809...35K} might lead to an increase in the energy content of nonthermal electrons and could explain the high values of energy in nonthermal electrons estimated. This is in part because of the use of the rather simple way in which $E^*_c$ is estimated; however, it is also because of the low values of the temperature $T$ ($\simeq 0.8$~keV) used by \citet{2016ApJ...832...27A} in applying this simple estimate. In deriving the approximate formula $E^*_c = \delta \times k_B T$, \citet{2015ApJ...809...35K} stress that the value of $T$ used must be the value of $T$ corresponding to the Maxwellian thermal plasma in the loop.

\begin{figure*}[t]\centering
\includegraphics[width=0.48\linewidth]{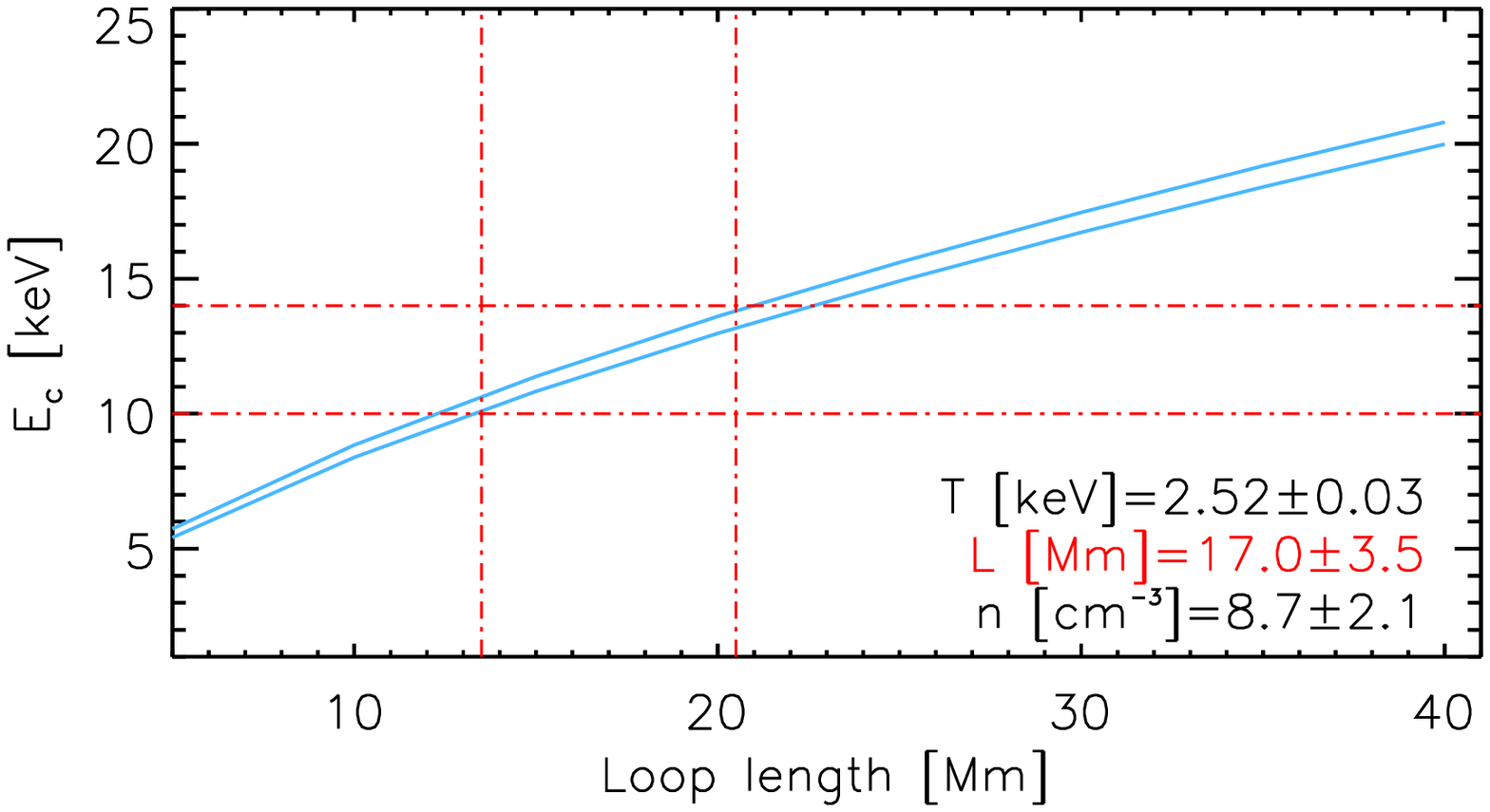}
\includegraphics[width=0.48\linewidth]{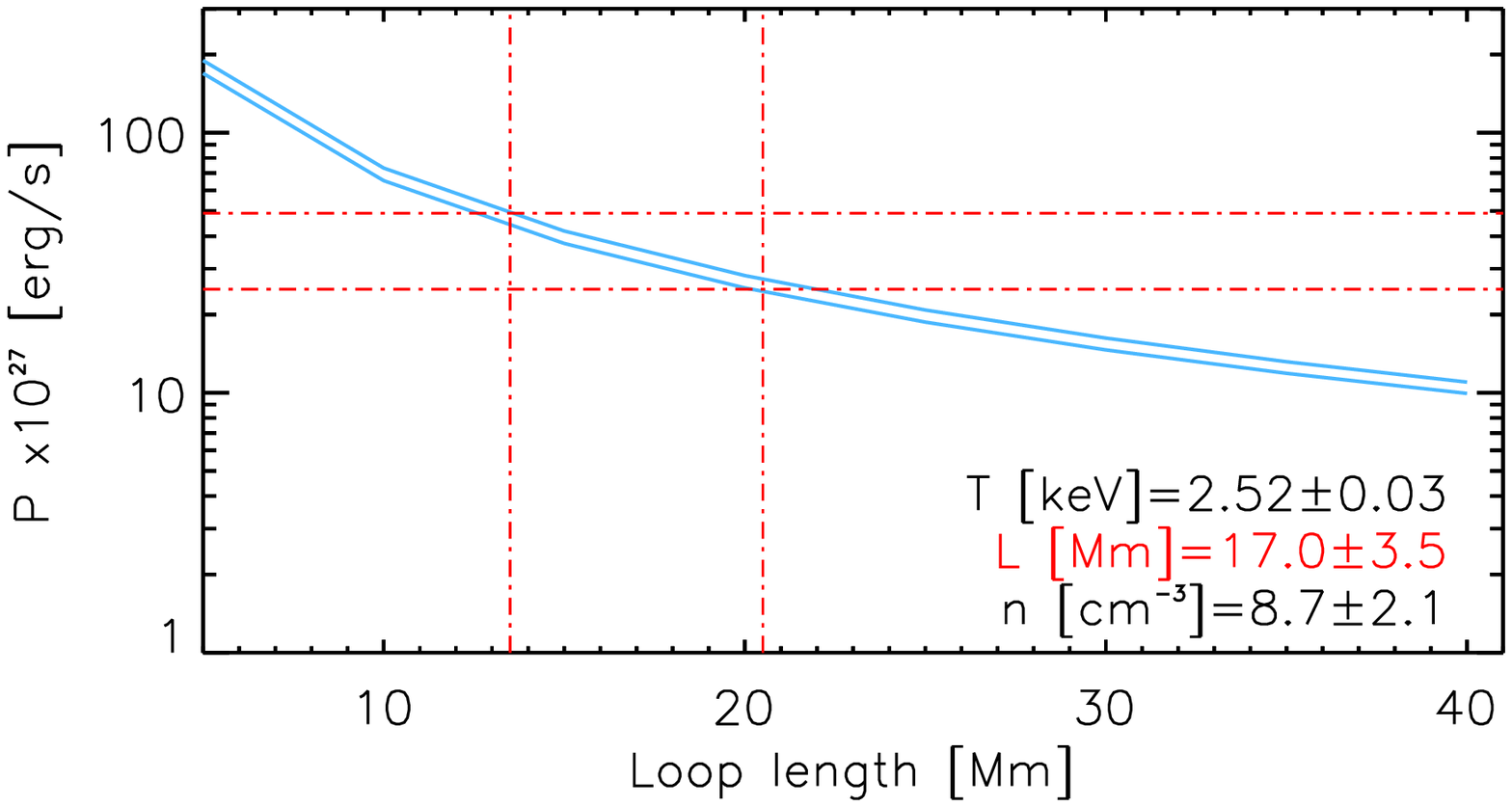}
\includegraphics[width=0.48\linewidth]{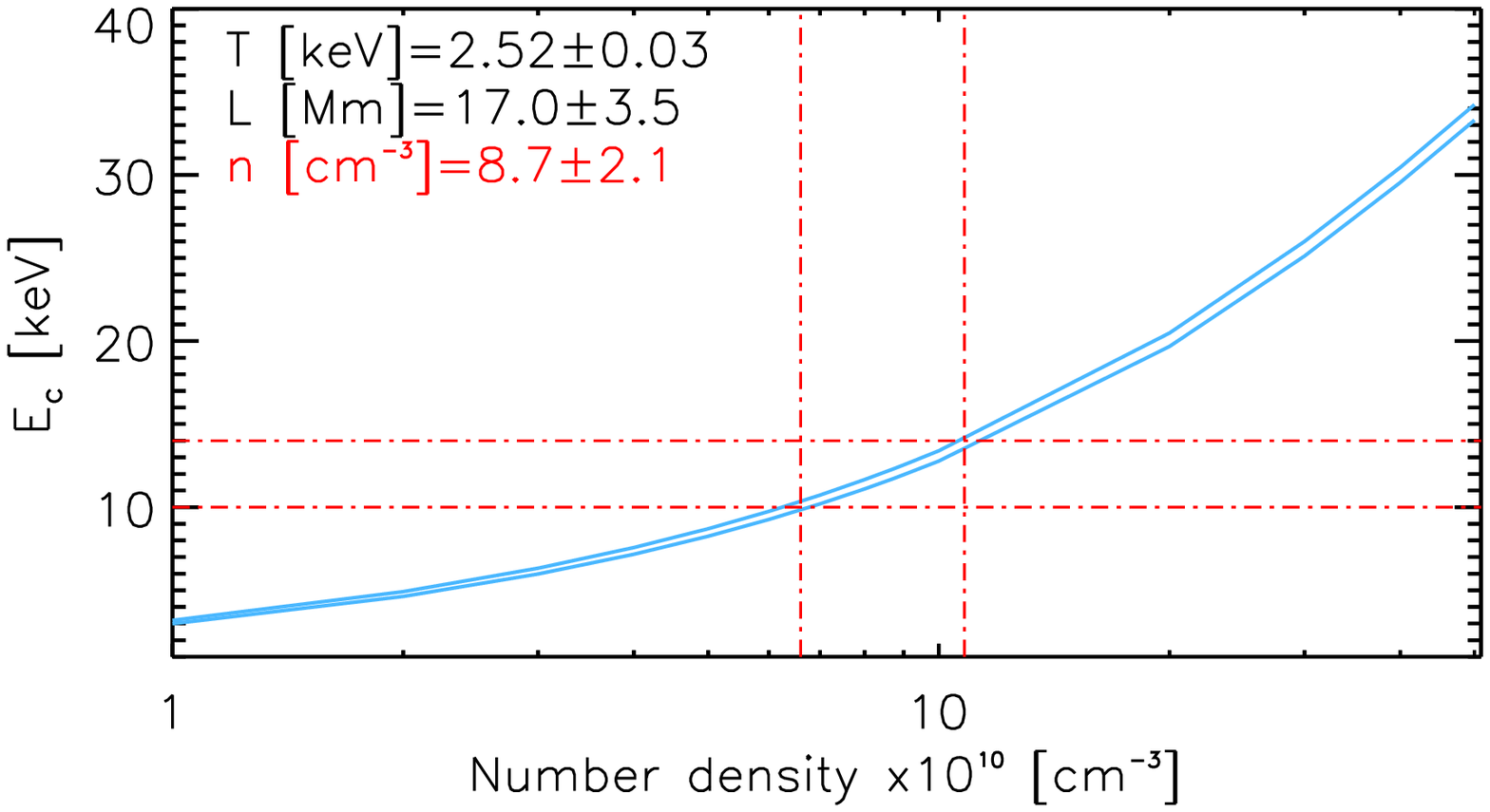}
\includegraphics[width=0.48\linewidth]{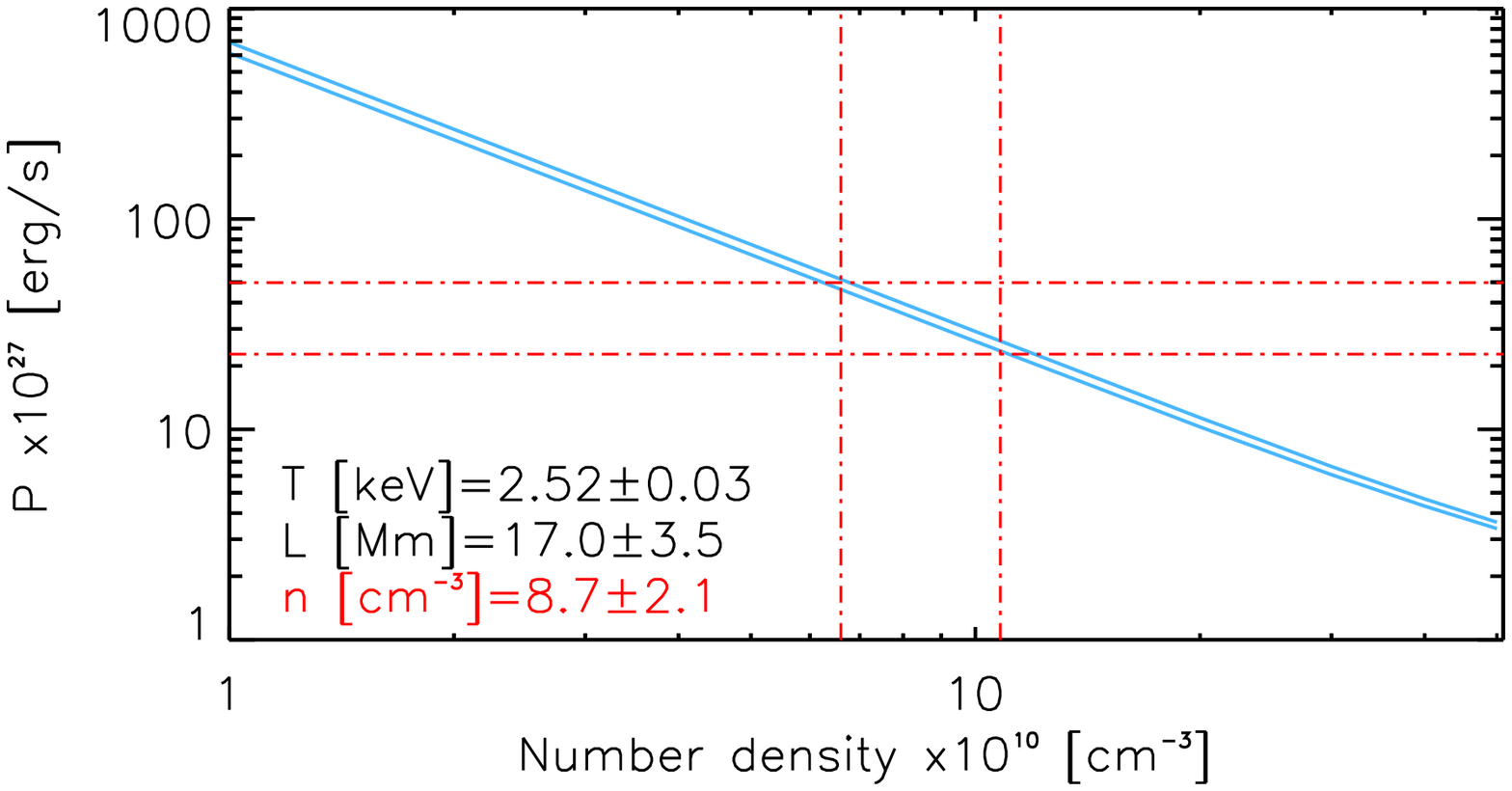}
\includegraphics[width=0.48\linewidth]{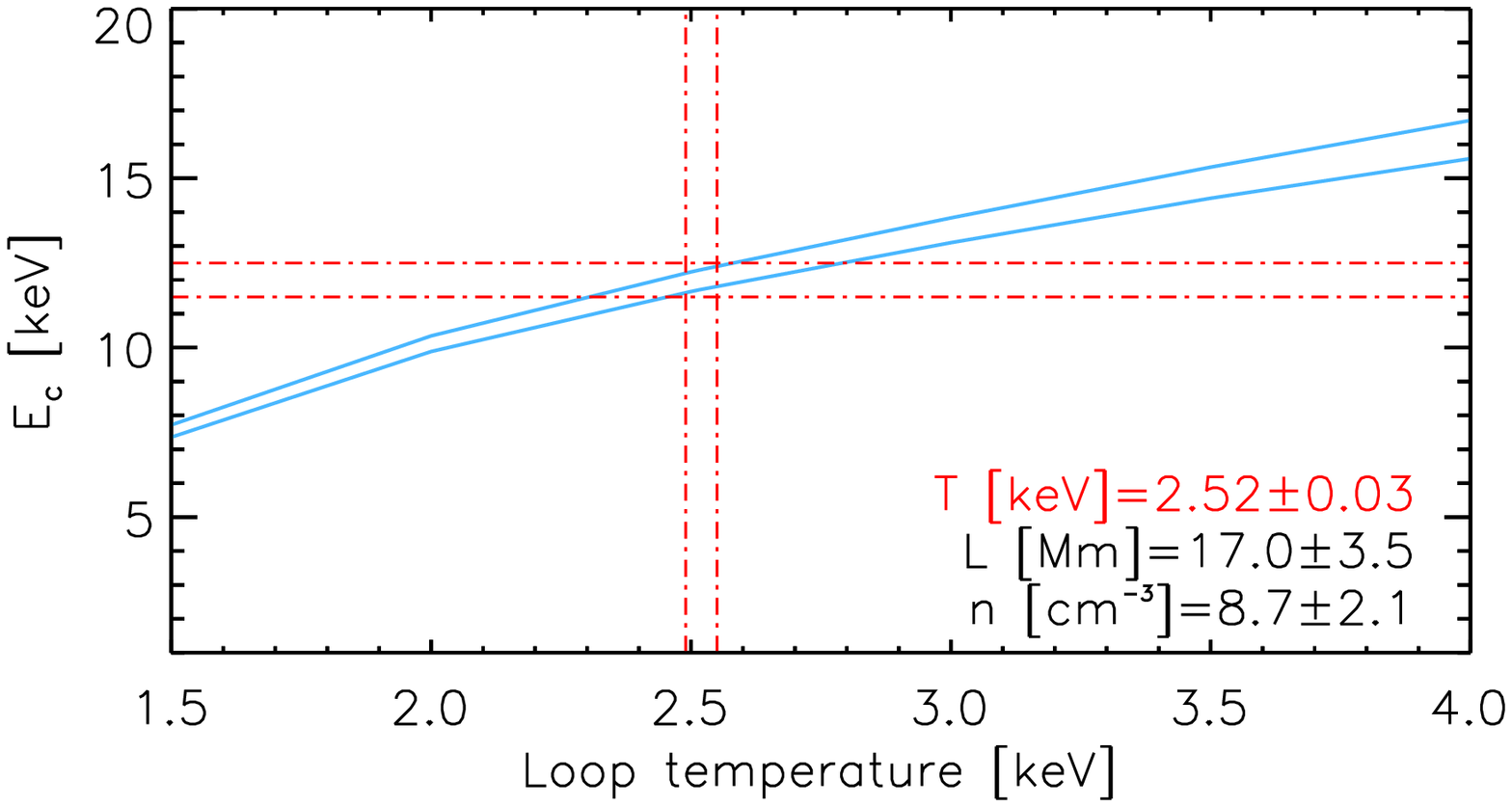}
\includegraphics[width=0.48\linewidth]{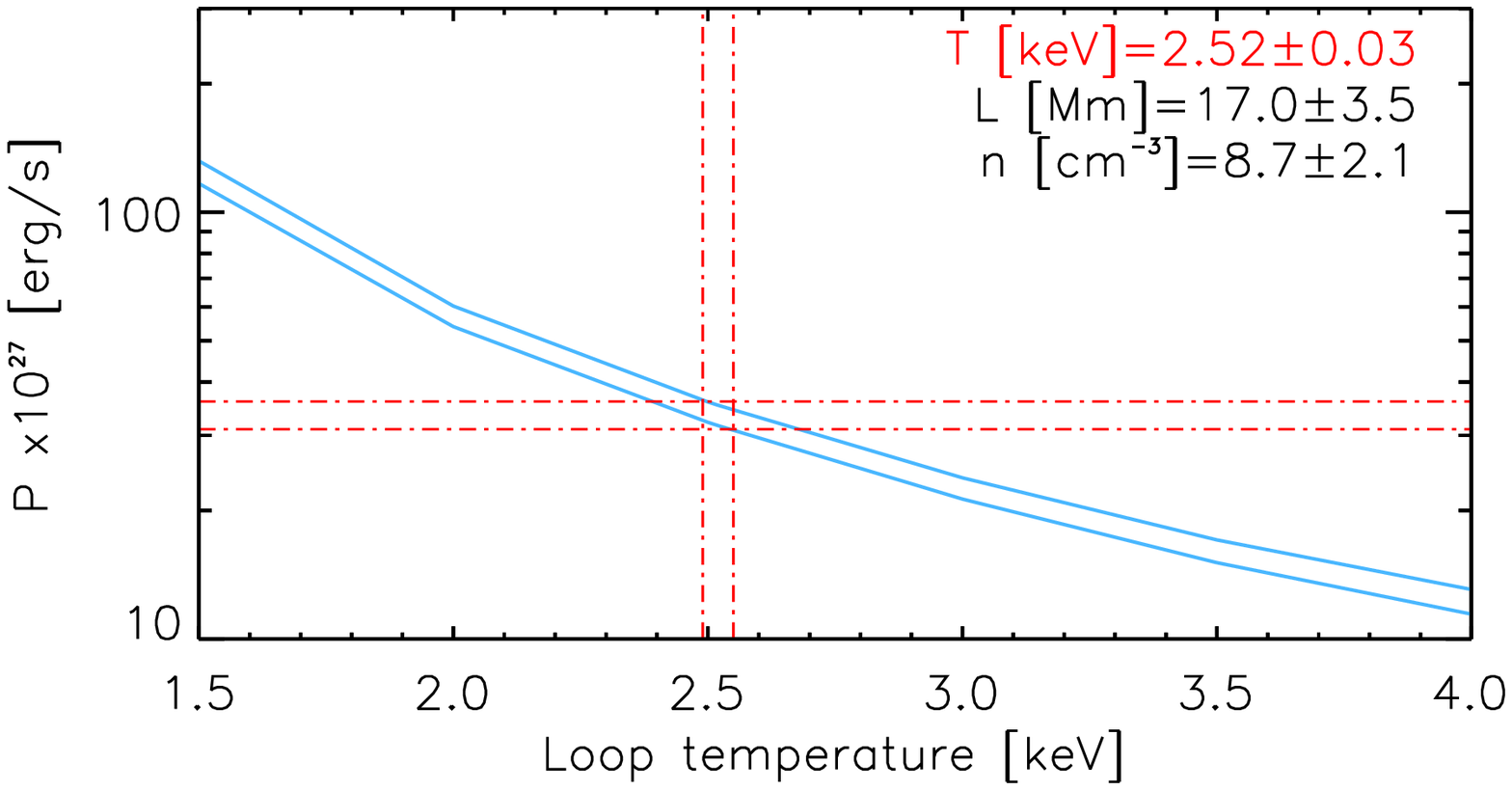}
\caption{An example of how the nonthermal electron parameters are constrained by the surrounding plasma environment in a warm-target. Low-energy cutoff $E_{c}$ {\em (left)} and nonthermal electron power $P$ {\em (right)} versus half loop length $L$~[Mm] {\em (top)}, loop number density $n_{\rm loop}$ [cm$^{-3}$] {\em (middle)} and loop temperature $T_{\rm loop}$ [keV] {\em (bottom)}. In this example, the values used to constrain the nonthermal power of SOL2013-05-13T02:12 are shown by the vertical dashed-dot red lines ($\pm$ uncertainty - dashed red lines), and in Table \ref{tab1}. The horizontal dashed-dot red lines show the resulting uncertainties in $E_{c}$ and $P$. Parameters determined from imaging (i.e $n_{\rm loop}$ and $V$) lead to an additional uncertainty in $E_{c} \pm \sim 2$~keV, not accounted for in OSPEX.}
\label{fig 4}
\end{figure*}

\subsection{Influence of the thermal parameters on the inferred value of the electron power}\label{plasma}

Of the three warm-target parameters $n_{\rm loop}$, $T_{\rm loop}$ and $L$, only $T_{\rm loop}$ can be obtained solely from the HXR spectrum; the other two parameters $n_{\rm loop}$ and $L$ can only be obtained by adding information from X-ray imaging.  It is therefore instructive to investigate how variations in each of the thermal parameters change the resulting values of the nonthermal electron parameters, in particular $E_{c}$ and $P$.

In the three sets of panels in Figure~\ref{fig 4}, we plot the values of $E_{c}$ and $P$. The top two panels show the variations with $L$ for fixed values of $T_{\rm loop}$ and $n_{\rm loop}$; the middle panels show the variations with $n_{\rm loop}$ for fixed values of $T_{\rm loop}$ and $L$, and the bottom two panels show the variations with $T_{\rm loop}$ for fixed values of $n_{\rm loop}$ and $L$.  We note the following behaviors:

\begin{itemize}

\item As $L$ becomes smaller (decreasing from $40$~Mm to $5$~Mm), $E_{c}$ decreases from $\simeq$20~keV to $\simeq$5~keV and the power $P$ correspondingly increases.  This is because the smaller column density in the corona means that fewer accelerated electrons thermalize there. The estimated uncertainty in $L$ of $\pm3.5$~Mm leads to an uncertainty of $\pm2$~keV in $E_{c}$ that is not accounted for in OSPEX.

\item Decreasing the plasma number density $n_{\rm loop}$ similarly lowers the coronal column density and hence the number of accelerated electrons thermalized there; a similar variation in $E_{c}$ hence results. Again, the estimated uncertainty in $n_{\rm loop}$ of $\pm2.1$~cm$^{-3}$ leads to an uncertainty of $\pm2$~keV in $E_{c}$ that is not accounted for in OSPEX.

\item Keeping the temperature $T$  from f\_vth fixed but varying the f\_thick\_warm parameter $T_{\rm loop}$ ($\ne T$) between $T_{\rm loop} = 4.0$~keV and $T_{\rm loop} = 1.5$~keV decreases $E_{c}$ from $\sim$16~keV down to $\sim$7~keV.  This is because, once again, a lower value of $T_{\rm loop}$ results in a smaller number of accelerated electrons blending into the thermal background, so that the nonthermal tail now extends to lower values of $E_c$.

\end{itemize}

We note that an isothermal fit was used to determine the plasma parameters in the flaring corona, even though it has been shown by \citet{2015A&A...584A..89J}, using a combination of X-ray spectroscopy and imaging, that for SOL2013-05-13T02:12 the coronal source cannot be strictly isothermal.  Therefore, the plasma temperature should be viewed as an average temperature throughout the volume in which the electrons propagate. The number density $n_{\rm loop}$ should also be viewed as an average value.

In the above method, we determined EM$_{0}$ from a time interval before the interval studied, but this approach is not the only way to determine the thermal properties of the corona. As an illustrative example, another approach, motivated by the fact that the plasma parameters change as the flare evolves (i.e. chromospheric evaporation, conduction, nonthermal electron heating), is to estimate the plasma parameters using the average values inferred from time interval both before (i.e., 02:08:52-02:09:00~UT) and after (i.e., 02:10:00-02:10:08~UT) the studied interval. This gives average thermal properties of EM$_{0}=(7.37\pm0.45) \times 10^{48}$~cm$^{-3}$,  $T=(2.52\pm0.04)$~keV, and $n_{\rm loop}=(9.3\pm2.2) \times 10^{10}$~cm$^{-3}$. Using these values in the main interval of 02:09:00-02:10:00~UT gives electron parameters of $\delta=4.14\pm0.03$, a significantly larger $E_{c}=(29.0 \pm 1.9)$~keV, a commensurately smaller value of $\dot{N_0}=(0.7\pm0.1) \times 10^{35}$~s$^{-1}$. The inferred electron power is now $P=(4.8\pm0.8)\times10^{27}$~erg~s$^{-1}$. Thus, this is an excellent example that demonstrates how important it is to determine the thermal properties of the corona as accurately as possible with future X-ray instrumentation, so that the nonthermal electron parameters are determined as accurately as possible using the warm-target fitting method provided.

\section{Summary and discussion}\label{discussion}

The procedure presented in this paper allows us to infer from HXR spectra the characteristics of accelerated electrons, including the electron acceleration rate and the nonthermal electron power. This is achieved by self-consistently considering the evolution of near-thermal electrons, the electrons that carry the bulk of the energy in solar flares. Previous flare energy estimates based on a cold thick-target model provide only an upper limit on the cutoff energy $E_c$ and hence a lower limit on the power $P$ and the energy contained in nonthermal electrons. Furthermore, the low-energy cutoff $E_{c}$ is very poorly constrained by the cold thick-target model, so that the nonthermal electron power $P$ (Figure \ref{fig1}, bottom panel) can range over two orders of magnitude.

By contrast, our warm-target fitting procedure (available as f\_thick\_warm.pro in OSPEX) provides not only an acceptable fit in terms of $\chi ^2$ (reduced $\chi^{2}\approx1$), but provides estimates of the electron power $P$ that lie within a range of $\pm (6-25)\%$ at the $1\sigma$ level, as shown in Figure \ref{fig1}. The electron acceleration rate (s$^{-1}$) and power (erg~s$^{-1}$) are now well constrained because the warm-target approach preserves the number of electrons, which all contribute to various parts of the observed X-ray spectrum.  An important feature of the warm-target model is that the density and temperature of the flaring loop are essential inputs to the model and therefore future instruments \citep[such as FOXSI,][]{2017AGUFMSH44A..07C} should aim to determine the thermal properties of the corona with as high a temporal and spatial resolution as possible. The dependency of the electron power $P$ on these plasma parameters opens a new avenue of research as to whether the temperatures and densities derived from {\em RHESSI} data are representative of the conditions at the field lines about which the electrons spiral, and allows the user to determine and constrain the electron properties in different plasma conditions and flare scenarios.

The self-consistent consideration of both thermal and nonthermal electrons provides more realistic estimates for the energy contained in accelerated electrons in solar flares. The detailed procedure explained herein (and the availability of the f\_thick\_warm.pro function in OSPEX) enables such calculations to be straightforwardly made. While the methodology based on  the cold target model cannot meaningfully constrain the low-energy cutoff and hence determine the energy in accelerated electrons with acceptable precision, the use of the warm-target procedure presented here {\it does} constrain the low-energy cutoff to within $\sim (7-14)$\% uncertainty at the $3\sigma$ confidence level; this allows the quantitative study of the contribution of accelerated electrons to overall solar flare energetics with uncertainties less than or comparable to those associated with other energetic components.

\acknowledgments The authors are thankful to the referee for insightful comments and to Alexander Warmuth for useful discussions. EPK and NLSJ were supported by a STFC consolidated grant ST/P000533/1. AGE was supported by grant NNX17AI16G from NASA's Heliophysics Supporting Research program.

\appendix

\section{Kappa-function injected spectrum}\label{kappa_section}

\begin{figure}[pht]\centering
\includegraphics[width=0.8\linewidth]{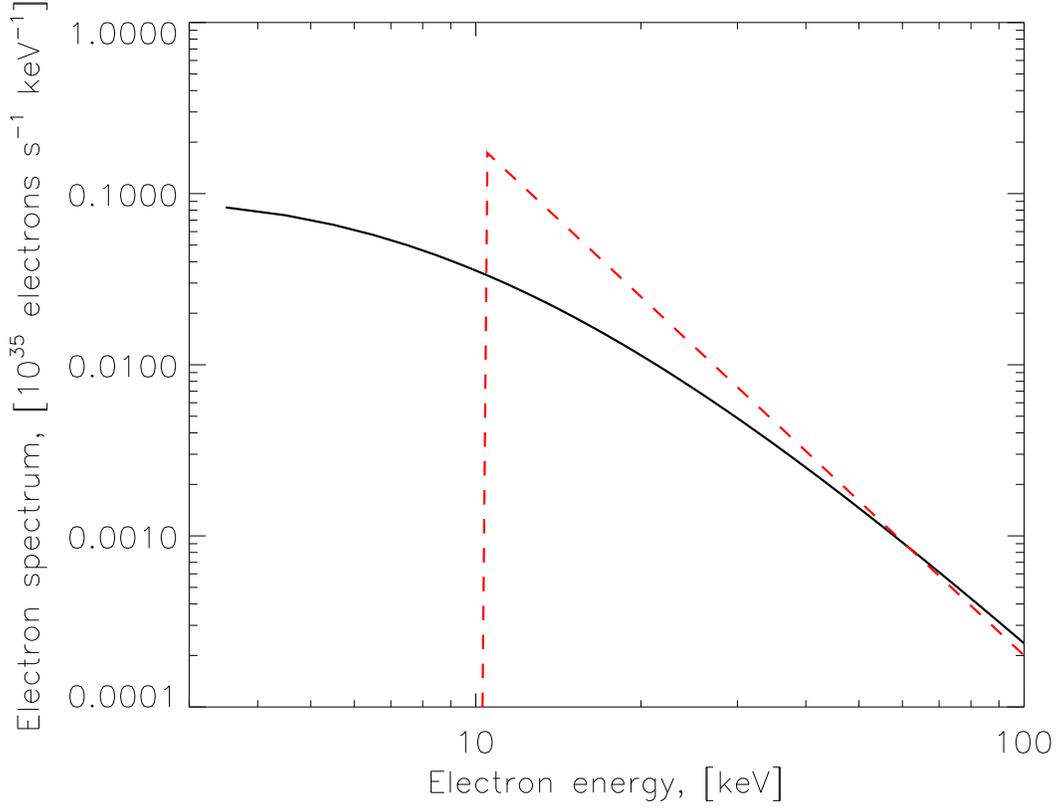}
\caption{Electron injection spectrum $\dot{N}(E)$ for a kappa distribution with $\kappa=4$, $k_{B}T=1.8$~keV (Equation~(\ref{eq:kappa_Norm}); solid line) and a power-law with $\delta =3$ and a 10~keV low-energy cutoff (Equation~(\ref{eq:tt_ndot}); red dashed line).}
\label{fig5}
\end{figure}

\citet{2014ApJ...796..142B} have demonstrated that the stochastic acceleration of electrons in the presence of Coulomb collisions leads naturally to an accelerated electron distribution (electrons~cm$^{-3}$~(cm~s$^{-1}$)$^{-3}$) that takes the form of a kappa-distribution (their Equation~(17))

\begin{equation}\label{eq:kappa_f}
f_\kappa(v)=
\frac{n_{\kappa}}{\pi^{3/2} \, v_{\rm te}^{3} \, \kappa^{3/2}} \, \frac{\Gamma(\kappa)}{\Gamma \left (\kappa-\frac{3}{2} \right ) } \,
\left ( 1+\frac{v^{2}}{\kappa \, v_{\rm te}^{2}} \right )^{-\kappa} \,\,\, ,
\end{equation}
where $v_{\rm te} = \sqrt{2 k_B T_\kappa/m_e}$ is the thermal speed and $\kappa = \Gamma_{\rm coll}/2 D_0$ is $\frac{1}{2} \times$ the ratio of the collisional and diffusional coefficients in the Fokker-Planck equation describing electron transport.  Equation~(\ref{eq:kappa_f}) is valid at all velocities; there is no ``low-energy cutoff'' in the accelerated electron spectrum.  For a collision-dominated environment, $\kappa$ is large and the distribution~(\ref{eq:kappa_f}) approximates a Maxwellian form $\exp(-v^2/v_{\rm te}^2$), while for a turbulence-dominated environment, $\kappa$ is small and the distribution~(\ref{eq:kappa_f}) approximates a power-law form $\sim$$v^{-2 \kappa}$. Figure \ref{fig5} demonstrates the similarities and the differences between two spectra from Equation (\ref{eq:delta_EM_kappa}) and from Equation (\ref{eq:fbar_warm2part}). At high energies $E\gg k_BT$, the distributions look like a power-law $\dot{N}_\kappa(E)\propto E^{-\kappa+1}$, but they behave differently in the deka-keV range (Figure \ref{fig5}).

The accelerated electron spectrum $\dot{N}_\kappa(E)$ corresponding to the form~(\ref{eq:kappa_f}) is

\begin{equation}\label{eq:kappa_N}
\dot{N}_\kappa(E) = A \, v \, f_\kappa(v) \, 4\pi v^2 \, \frac{dv}{dE}= A \, n_\kappa \, \left ( \frac{8}{\pi m_e k_B T_\kappa} \right )^{1/2} \frac{ \Gamma (\kappa) }{\kappa^{3/2} \, \Gamma (\kappa-\frac{3}{2})} \, \frac{E/k_BT_\kappa}{(1+E/\kappa k_B T_\kappa)^\kappa} \,\,\, ,
\end{equation}
where $A$ is the cross-sectional area of the loop. The total rate of electron injection is

\begin{equation}\label{electron-rate-kappa}
\dot{N_0} = \int_0^\infty \dot{N}_\kappa(E) \, dE = 2 \, A \, n_\kappa \, \left ( \frac{2 k_B T_\kappa}{m_e} \right )^{1/2} \frac{\kappa^{1/2}}{(\kappa-2) \, B(\kappa - \frac{3}{2}, \frac{1}{2}) } \,\,\, ,
\end{equation}
where $B(\cdot, \cdot)$ is the beta function. Equation~(\ref{eq:kappa_N}) can thus be written as

\begin{equation}\label{eq:kappa_Norm}
\dot{N}_\kappa(E) =  \left ( \frac{\dot{N_0} }{k_B T_\kappa} \right ) \,
\frac{(\kappa-1)(\kappa -2)}{\kappa^2} \, \frac{E/k_BT_\kappa}{(1+E/\kappa k_B T_\kappa)^\kappa} \,\,\, .
\end{equation}
Equation~(\ref{eq:kappa_Norm}), together with Equation~(\ref{eq:fbar_warm2part}), gives

\begin{equation}\label{eq:fbar_warm2part_kapp}
 \begin{split}
 \langle nVF \rangle (E) & \simeq
\Delta EM_\kappa \, \sqrt{\frac{8}{\pi m_e}} \, \frac{E}{(k_BT)^{3/2}} \, e^{-E/k_B T } + \frac{E}{K}\int_{E}^\infty \dot{N}_\kappa(E_0) \, dE_0 \,\,\, =\\
&=\Delta EM_\kappa \, \sqrt{\frac{8}{\pi m_e}} \, \frac{E}{(k_BT)^{3/2}} \, e^{-E/k_B T } +
\dot{N_0} \, \frac{E}{K} \,
\frac{1 + \left (1 - \frac{1}{\kappa} \right ) \frac{E}{k_B T_\kappa} }{ ( 1 + E/\kappa k_B T_\kappa )^{\kappa-1}} \,\,\, ,
 \end{split}
\end{equation}
where, using Equations~(\ref{eq:E_min_adj}) and~(\ref{eq:delta_EM}) and the fact that the kappa distribution is flat at low energies,

\begin{equation}\label{eq:delta_EM_kappa}
\Delta EM_\kappa  \simeq
\frac{\pi}{K} \, \sqrt{\frac{m_e}{8}} \, (k_B T)^2 \, \frac{\dot{N_0}}{E_{\rm min}^{1/2}}
\simeq \frac{\pi}{K} \, \sqrt{\frac{m_e}{8}} \, \frac{(k_B T)^2}{E_{\rm min}^{1/2}} \, \dot{N_0} \simeq
\frac{\pi}{K} \, \sqrt{\frac{m_e}{24}} \, \left ( \frac{L}{5 \lambda} \right )^2 \, (k_B T)^{3/2} \, \dot{N_0}  \,\,\, ,
\end{equation}
where $\dot{N_0}$ (s$^{-1}$) is the total injected rate. Equations\footnote{Equation (\ref{eq:delta_EM_kappa}) is now included in OSPEX as f\_thick\_warm\_kappa.pro.} (\ref{eq:fbar_warm2part_kapp}) and~(\ref{eq:delta_EM_kappa}) (cf. Equations~(\ref{eq:fbar_warm2part}) and~(\ref{eq:delta_EM})) give the mean electron flux corresponding to the injection of a kappa distribution, and it should be re-emphasized that both the thermal and nonthermal parts of Equation~(\ref{eq:fbar_warm2part_kapp}) are determined by the form of ${\dot N}(E_0)$.

\bibliographystyle{aasjournal}
\bibliography{refs}

\end{document}